# Evaluation of Non-Invasive Thermal Imaging for detection of Viability of Onchocerciasis worms

Ronak Dedhiya[1], Siva Teja Kakileti[1], Goutham Deepu[1], Kanchana Gopinath[1], Nicholas Opoku[2], Christopher King[3], Geetha Manjunath[1]

*Abstract*— Onchocerciasis is causing blindness in over half a million people in the world today. Drug development for the disease is crippled as there is no way of measuring effectiveness of the drug without an invasive procedure. Drug efficacy measurement through assessment of viability of onchocerca worms requires the patients to undergo nodulectomy which is invasive, expensive, time-consuming, skill-dependent, infrastructure dependent and lengthy process. In this paper, we discuss the first-ever study that proposes use of machine learning over thermal imaging to non-invasively and accurately predict the viability of worms. The key contributions of the paper are (i) a unique thermal imaging protocol along with pre-processing steps such as alignment, registration and segmentation to extract interpretable features (ii) extraction of relevant semantic features (iii) development of accurate classifiers for detecting the existence of viable worms in a nodule. When tested on a prospective test data of 30 participants with 48 palpable nodules, we achieved an Area Under the Curve (AUC) of 0.85.

*Clinical Relevance*— This is the first ever research effort of using thermal imaging in the assessment of viability of onchocerca worms and it resulted in a very high specificity>95% which makes it a promising modality to pursue further.

I. INTRODUCTION

Onchocerciasis is the world's second-leading infectious cause of blindness. Over a half a million people are blind or visually impaired due to this disease. It is caused by the filarial worm called Onchocerca volvulus that is transmitted through repeated bites of infected blackflies. During a blood meal, an infected blackfly introduces larvae onto the skin of the human host. In subcutaneous tissues, the larvae develop into adult filariae, which commonly reside in nodules in subcutaneous connective tissues [1]. The control in the spread of the disease and its elimination has been a high priority in Ghana and worldwide which are based on the annual or biannual treatment with ivermectin. Ivermectin temporarily blocks transmission of infection by clearing microfilariae (eggs) but does not kill the adult worms [2]. Unfortunately, there are currently no safe and easily administered drug(s) that kill and/or sterilize adult worms which can live up to 15 years.

For delivering treatment effectively, it is important to know if the adult worms are alive or dead (in other words 'viability'). Currently, the only way to determine this is by surgical removal of subcutaneous nodules and subsequent histological examination, which is expensive, time-consuming, skill-dependent, infrastructure dependent and lengthy process [3]. An alternate technique used for diagnosis is analysis of skin biopsies, also known as skin snips, by microscopy or molecular techniques. It measures microfilariae density in the skin tissue but is relatively less accurate in classification of viability of the worms [4]. It also has low throughput and can be painful for the patient. Therefore, a significant bottleneck in facilitating drug development and testing efficacy of drug is the unmet need of a non-invasive method that identifies the viability of worms.

A non-invasive technique like thermal imaging has been used to study breast health and other conditions based on variations in skin temperature due to blood flow variations, changes in metabolic activities of tissue and presence of infections [5-6]. Some recent studies show its high potential in the diagnosis of breast cancer, diabetes neuropathy and peripheral vascular disorders [5-6]. Since the onchocerca worms are found to reside in subcutaneous tissue and are observed to have high metabolic activity due to angiogenesis [7], our thesis was that we could use thermal imaging for detecting the viability of onchocerca worms by detecting thermal signature due to metabolic activity of onchocerca worms that traversed to the skin surface. The latest thermal cameras can measure minute variations up to 0.05 °C and could pick up these thermal variations and help in the assessment of the viability of onchocerca worms. However, there is no prior work to support or contradict this hypothesis. In this paper, we discuss a novel pilot study exploring the use of thermal imaging coupled with machine learning for detection of viability of onchocerca worms. As there is no prior knowledge on the visual interpretation of thermal images for onchocerca viability, there is no prior medical knowledge to assess thermal data with visual analysis. We wanted to explore computer aided analysis with machine learning to derive this knowledge and assess its effectiveness especially when we have little or no knowledge on the relation between input and output. In section 3 of this paper, we discuss a machine learning methodology that we used to study the thermal variations of the onchocerca worms to differentiate their viability. Section 4 describes the experimentation and results that shows the potential of thermal imaging to be a promising tool for onchocerca viability detection.

Research supported by Bill & Melinda Gates Foundation.,
[1]Ronak, Dr. Siva Teja, Goutham (Intern), Dr. Kanchana and Dr. Geetha M are from NIRAMAI Health Analytics Pvt. Ltd, INDIA. (Contact email: geetha@niramai.com).
[2]Dr. Nicholas Opoku is from UHAS Ghana.
[3]Dr. Christopher King is from CWRU USA.

## II. DATASET DESCRIPTION

### A. Study Site & Population

A pilot study for evaluating onchocerca viability with thermal imaging was conducted at Hohoe, Ghana during July 2020 to December 2021. Clearance was granted by the ethics committee of the UHAS-REC (University of Health & Allied Science Research Ethics Committee), GHS-REC (Ghana Health Service Ethics Review Committee) and CWRU-IRB (Case Western Reserve University Institutional Review Board).

125 participants with presence of one or more accessible, visible and palpable onchocerca nodules were recruited for the study. Their ages ranged from 18 to 70 years with a mean of 42 years. After taking informed consent, participants underwent routine physical examination followed by thermal Imaging, skin snip and nodulectomy. The histology results of nodulectomy were considered as ground truth to train and validate the ML models. In total, 192 nodule locations were identified from 125 participants. Nodulectomy showed 101 of these nodule locations had one or more alive female worm and the remaining 91 nodule locations had no worms or dead worms.

### B. Thermal Image Acquisition

Fig. 1 illustrates the procedure followed for thermal data acquisition that we developed. Firstly, technician identified palpable nodules for each participant and marked them with a marker on the body which are then used for nodule registration. Participants were then relaxed for 5 minutes under room temperature to allow extraneous heat to dissipate, and a thermal image of the nodule site was captured. An additional localized cooling using alcohol swabs was applied to the nodule site and a thermal video of 120 seconds duration was recorded. The localized cooling was applied on and around the nodule site up to 5 cm radius. Localized cooling aids in enhancing the thermal contrast between a nodule and healthy skin due to their differences in thermal recovery process. FLIR E75 infrared camera with a thermal resolution of 320x240, a thermal sensitivity of 40 mK and a frame rate of 30 Hz was then used to capture the images and videos.

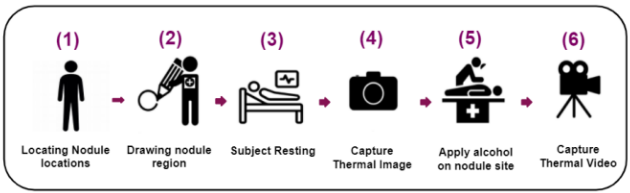

Figure 1. Thermal acquisition protocol followed in the study.

## III. METHODOLOGY

To assess the viability of onchocerca nodules, as it was a small data set, hand-crafted semantic features were extracted from both thermal video and thermal image. However, before the extraction of features, a set of pre-processing steps are required as illustrated in Fig. 2 and described below.

### A. Alignment of Image snapshot and thermal video

For each participant, an image before localized cooling and a video after localized are captured as a part of protocol. There is often a misalignment between this captured image and video due to their difference in timings during image capture. Also, the frames in the video are also mis-aligned due to hand movement or involuntary movement of the participant. Though the use of tripod might solve the issue of hand movement, it is difficult to image certain regions such as Sacrum, illiac crest etc. In order to study the heat variations over time, it is important to register both the image and the frames in the video. In the literature, there are several image registration techniques to align the frames in the video. We found that Enhanced Correlation Coefficient (ECC) [8] based geometric transformation performs better for thermal images. ECC uses iterative non-linear optimization to estimate geometric transformation which increase the correlation between frames. The first frame of the video is used as a reference image and all the frames of the video and the image before localized cooling are aligned with respect to the initial frame. This allows us to track temperature changes at a point or a region of interest throughout the video sequence and measure the effect of localized cooling by comparing the temperature values of video frames with the image before cooling.

### B. Nodule Registration across thermal and visual images

As thermal images capture additional skin region apart from the nodule region, it is important to know the actual location of nodule to study the thermal patterns of onchocerca nodules. To obtain this location information, the nodule region is marked on the participant's body before image capture as a part of imaging protocol. These regions are then marked manually on the thermal image that is captured before cooling with the assistance of the corresponding visual image that is captured by thermal camera. ECC transformation that is described above is applied to this nodule location as well to perform the nodule registration.

### C. ROI Segmentation

Another pre-processing step that is required for feature extraction is the segmentation of region of interest. The captured thermal image/video consists of background region like cloth, walls etc. along with alcohol applied region of interest. It is important to segment the alcohol applied region to study the healthy tissue and nodule sites. For this purpose, we employed a 3-stage encoder-decoder architecture called V-Net [9] to automatically segment the region of interest belonging to alcohol applied region. Encoder stages down-samples the input thermal image to extract meaningful information, which is then used by decoder stages to reconstruct the segmentation output indicating the region of interest (alcohol applied region). The ROI segmentation is applied for initial frame of the video and the same segmentation mask is used to get the ROI mask for the remaining frames and the image as they are already aligned.

### D. Feature Extraction

To differentiate the viability of onchocerca worms, we broadly extracted five sets of features as follows.

*i. Temporal Features:* These features characterize the thermal variations over time. Thermal patterns of Onchocerca nodules are supported by additional blood supply and resources for the growth of worms [7]. The heat pattern released from this increased activity is controlled more by the onchocerca nodule than the sympathetic nervous system. This might lead to differences in thermal variation of the onchocerca nodule region compared to a healthy site. To capture this, we extracted 42 time series features such as AUC, slope, skewness, kurtosis, spectral centroid, spectral slope and frequency using the mean and standard deviation time graph for entire alcohol region, 20x20 and a 40x40 windows around nodule region.

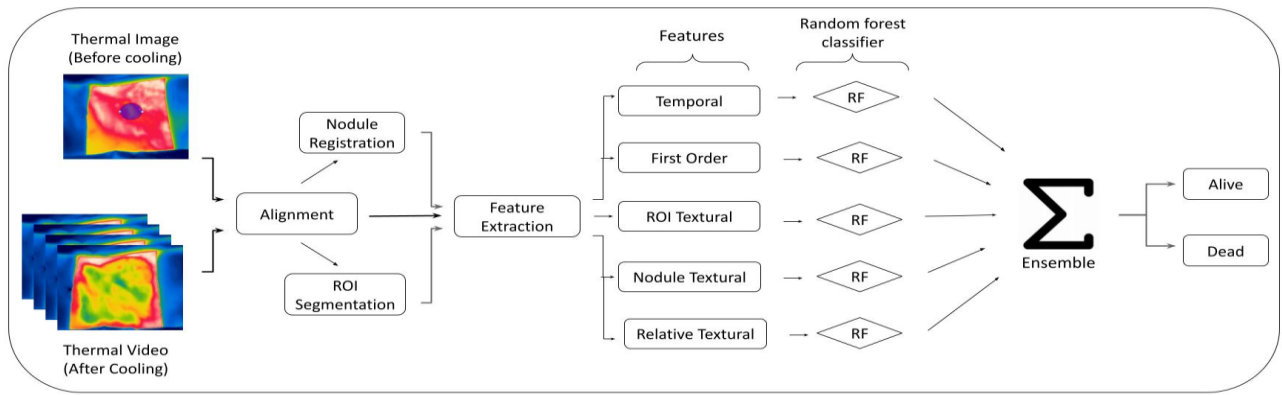

Figure 2. Block Diagram representing machine learning (ML) pipeline for onchocerca viability detection.

*ii. ROI Textural Features:* Textural features characterize the spatial distribution of a region. Since the metabolic activity of nodule affects the surface heat pattern, the texture of the captured thermal image might be different from healthy tissue. For this, second order statistical texture features namely contrast, dissimilarity, homogeneity, energy, correlation and angular second moment are extracted from gray level co-occurrence matrix (GLCM) [10] of the ROI. A total of 576 features were obtained when GLCM is computed at distances = {1, 3, 5} and angles = {0, 45, 90, 135} for the uncooled image and aligned video frames at every 15 seconds.

*iii. Nodule Textural Features:* These features capture the texture of nodule region alone. Similar to above, 576 feature are extracted by considering only the nodule region instead of the entire ROI region.

*iv. Relative Textural Features:* This captures the relative texture between the ROI and nodule by computing the differences and ratios between the ROI and nodule textual features, which resulted in 2*576 features.

*v. First order Features*: These features capture the temperature histogram distribution of nodule and surrounding region at image/frame level. It includes first order temperature features such as minimum, mean, maximum, standard deviation and mode that are calculated for the ROI and nodule regions of the video frames at t = {0, 60, 120$^{th}$ sec}. To measure the relative differences between the nodules and healthy skin and as well as between the video frames, we also considered their differences. In total, 90 features are computed.

*E. Feature Selection*

As this is first ever pilot study, the collected dataset is very small when compared to the number of features extracted. Therefore, we reduce the dimensionality of these features by using Principal Component Analysis (PCA) [11]. The number of components for each of the above-described feature set is selected such that they produce at least 95% data variation of the training set. This is an important step before classification as the above sets of features have high correlation between the features.

*F. Classification*

Firstly, five separate random forest (RF) classifiers are trained using the five different feature sets to independently predict the viability of the onchocerca worms. RF is selected after comparing the performance with other classifiers such as logistic regression, support vector machines and multi-layer perceptron. In general, RF is found be a good discriminator for medical datasets and the use of decision trees resemble the human decisioning mechanism. To obtain the final classification, a voting mechanism is employed by counting the individual classification votes ($C_i$) towards the presence of alive female worm as represented in eq. 1.

$$F = \sum_{i=1}^{5} C_i \qquad (1)$$

Here, $C_i$ represent votes {0,1} from i$^{th}$ classifier and F is the sum of the votes.

## IV. EXPERIMENTS & RESULTS

As a part of this pilot study, the initial 95 participants' data comprising 79 live nodules and 65 dead nodule sites were split into training and validation data in a ratio of 80:20 (stratified). The remaining 30 participants' data were used as blind test set and had 22 live female nodule sites and 26 dead nodule sites. Histology of the nodules was used to assess the viability of nodules. Thermal Imaging for all the nodule sites were performed by a trained technician in Ghana. Data was loaded to in-house software tool which uses Alignment module (III.A) to instantly generate a registered video and thermal image for each participant. This was possible for all subjects since Alignment module is an image processing technique and does not require any training. The result of this automatic alignment was visually assessed, and a reimaging was done when there was an indication of wide misalignment between the frames in the thermal video and image.

For ROI segmentation of alcohol applied region, the in-house tool supported manual segmentation of the region of interests using a polygon user interface. We obtained the ROIs for all the nodule sites. As discussed in III.C, V-Net deep learning architecture was used for automatically detect this ROI from thermal images. The V-net was trained with training set involving 115 nodules data and validated on 29 nodule's data. To increase the training data, the initial 150 frames (5 sec) of the thermal video for each nodule site was used. The segmentation in the initial frame of the video was a slightly easier problem as applied alcohol region is colder than surrounding skin region and the training data of 115*150 were sufficient to give good results. Binary cross entropy and Adam optimizer were considered as loss function and optimization technique, respectively, to update the weights of the network. When we tested this trained model on the 48 nodules thermal data, we obtained a very good Dice index of 0.92. The visual assessment of these results is very close to manual segmentation and this technique can be used to predict the initial segmentation region that can be adjusted by a technician. This approach can save time and speeds the image capture in future.

Automated viability of the onchocerca nodules is the main goal of this study. To achieve this, we trained 5 different random forest classifiers with different feature sets. For better learning and to reduce overfitting, we considered 40 decision trees and limited the depth to 3. All these five classifiers are trained on the training set consisting of 115 nodules data and validated on the 29 nodules data. Once trained, an operating point with specificity>95% and sensitivity>60% was selected based on the validation dataset. In case of no such operating point, we lowered the sensitivity to obtain the model that has best specificity and a moderate sensitivity. This criterion was chosen as per the target product profile requirement for onchocerciasis elimination laid by WHO [4]. The final classification was performed using an ensemble approach involving voting mechanism. Table A. shows the results of the different classifiers and skin snip.

|  | *Sensitivity* | *Specificity* | *AUC* |
|---|---|---|---|
| Skin Snip | 82 | 76 | - |
| Temporal | 36.4 | 96 | 0.60 |
| ROI Textural | 31.8 | 100 | 0.68 |
| Nodule Textural | 22.7 | 96 | 0.61 |
| Relative Textural | 18.2 | 92 | 0.64 |
| First Order | 18.2 | 90 | 0.57 |
| Ensemble (Voting) | **68** | **96** | **0.85** |

Table A. Results comparing different classification approaches.

As seen from the table, the individual classifiers trained with different domain knowledges obtained from different features resulted in very high specificity, but a very low sensitivity. However, when we combined their collective knowledge with ensemble approach, a high accuracy was seen. Overall, the final ensemble classification resulted in a high specificity of 96% with a sensitivity of 68% when we considered votes>=2 for assessing the viability. These results are better compared to skin snip (skin biopsy), which resulted in a low specificity of 76%.

## V. CONCLUSION

The results obtained in this pilot study shows the potential for combination of thermal imaging and machine learning in detecting the viability of alive female worms. For drug efficacy testing use case, the operating point was chosen to have high specificity though with moderate sensitivity. The use of thermal imaging makes the diagnosis a completely non-invasive, portable and affordable approach when compared to on-field diagnostic solutions such as skin snip that are invasive, costly and skill-dependent. This is a first ever pilot study exploring the feasibility of thermal imaging in detecting the viability onchocerca worms. Large scale studies should be planned to evaluate its effectiveness for clinical implementation.

## ACKNOWLEDGMENT

This work was supported, in whole or in part, by the Bill & Melinda Gates Foundation [INV-007341]. Under the grant conditions of the Foundation, a Creative Commons Attribution 4.0 Generic License has already been assigned to the Author Accepted Manuscript version that might arise from this submission. In particular, we thank Dr. Jonathan Arm for his guidance in the project. We thank technicians from UHAS Ghana for their support in data collection, imaging and ground truth assessment.